\documentstyle[aps,twocolumn,floats,epsf,psfig]{revtex} 

\title{Anomalous Drude Model}

\author{Hermann Schulz-Baldes,} 
\address{Laboratoire de Physique Quantique,
Universit\'e Paul-Sabatier, 
118, Route de Narbonne, 31062 Toulouse Cedex, France.}

\begin{document}

\maketitle

%%%%%%%%%%%%
\begin{abstract}

A generalization of the Drude model is studied. 
On the one hand, the free motion of the particles is allowed to be sub-
or superdiffusive; on the other hand, the distribution of the time
delay between collisions is allowed to have a long tail and even a
non-vanishing first moment. The collision averaged motion is either
regular diffusive or L\'evy-flight like. The anomalous diffusion
coefficients show complex scaling laws. The conductivity can be
calculated in the diffusive regime. The model is of interest for the
phenomenological study of electronic transport in  quasicrystals.

\vspace{.2cm}


\end{abstract}

%%%%%%%%%%%

\vspace{.5cm}

Drude's model \cite{Dru,MA} considers a gas of independent charged  classical
particles. Their ballistic motion is perturbed by collisions at random
times distributed according to the Poisson process of rare events. At
every collision, a particle exchanges impulsion and energy with a
scatterer in a random way. The scatterers are supposed to be at
thermal equilibrium and the temperature of this bath partially
determines the distribution of the scattering parameters. The
collisions allow the particles to dissipate impulsion and energy to
the bath, but they also force the collision averaged time evolution 
to be diffusive. Placing the system in an exterior electric field, one
obtains the celebrated Drude formula for the conductivity in this
diffusive regime.

The first modification of the Drude model considers the free motion of
particles between collisions. Let us suppose that it is characterized
by an exponent $\sigma$ taking values in the interval $[0,1]$ and an
anomalous speed $\vec{v}_\sigma$ such that

\begin{equation}
\label{eq-diffexpo}
\vec{x}(t)\;=\;\vec{x}(0)+\vec{v}_\sigma t^\sigma
\mbox{ , }
\end{equation}

\noindent where $\vec{x}(t)$ is the position in euclidien physical
space at time $t$ and $\vec{x}(0)$ the initial position at $t=0$. Of
course, such a motion cannot be deduced from a Hamiltonian in the
framework of classical mechanics, but it does mimic the anomalous
quantum diffusion of a particle in a disorder or quasiperiodic on-site
potential where quantum interferences play an important r\^ole
\cite{Sir}. This has been shown numerically for the Fibonacci and the
Harper Hamiltonian \cite{HA} as well as for two-dimensional models for
quasicrystals \cite{SPB}. Furthermore there is analytical evidence for
the Fibonacci Hamiltonian \cite{Pie}. If the Fermi level is in a
region of localized states, $\sigma=0$ and (\ref{eq-diffexpo})
describes the hopping from one localized state to another. For the
Anderson model, one expects diffusive quantum motion ($\sigma=1/2$) in
three dimensions and at low disorder if the Fermi energy is in the
band center.

The particles undergo inelastic collisions at random times $t_n$,
$n\in{\bf Z}$, with the {\sl moving disorder}, notably phonons and
other particles of the gas.  The {\sl frozen} potential such as
impurity or quasiperiodic potential  has already been taken into
account and leads to the anomalous free motion (\ref{eq-diffexpo}). The
direction of the velocity vector $\vec{v}_\sigma$ after collision is
supposed to be completely random; the distribution of its modulus shall
be given by the temperature $T$ of the bath composed by the scatterers.
Solely the mean  $C_{T,\sigma}$ of $|\vec{v}_\sigma|^2$ will be needed
below.  In the quantum mechanical framework, this constant can be
calculated explicitly \cite{futwork}.

Finally, Drude's choice of the exponential law for the distribution of
the time delay $s$ between collisions is replaced by any probability law,
in particular those with a long tail. For sake of concreteness, let us
consider the family of probability measures on $[0,\infty)$ given by

\begin{equation}
\label{eq-delaydist}
p_{\mu,\tau}(s)ds\;=\;\frac{\mu}{(1+{s}/{\tau})^{1+\mu}}\;
\frac{ds}{\tau}
\mbox{ , }
\end{equation}

\noindent with $\mu>0$ and $\tau>0$. While the exponent $\mu$
characterizes the tail of the distribution $p_{\mu,\tau}$, $\tau$
allows to vary the (inelastic mean) collision time, namely the first
moment of $p_{\mu,\tau}$  (whenever it exists, that is $\mu>1$).
Phenomenologically, increasing the collision time corresponds to
lowering the temperature of the bath. Considering the number of
collisions up to a given time $t$ gives a continuous time stochastic
process with integer values, a so-called counting process. The only
counting process which is stationary and has independent increments
is the Poisson process \cite{Bau}. Hence the counting processes
defined by (\ref{eq-delaydist}) do not have this Markov property. Note
that, in particular, the probability to have no collision after time
$s+s'$ is bigger than the product of the probabilities to have none
after time $s$ and none after time $s'$. Let us finally remark that
the explicit form of $p_{\mu,\tau}$ is of no importance for the
results below as long as
$\limsup_{s\rightarrow\infty}s^{1+\mu+\epsilon}p_{\mu,\tau}(s)$ is
bounded for $\epsilon\leq 0$ and unbounded for $\epsilon>0$. Possible
physical reasons for the choice (\ref{eq-delaydist}) are not discussed
here.

\vspace{.2cm}

This accomplishes the presentation and motivation of the anomalous
Drude model characterized by the parameters $\mu,\tau,\sigma$ and $T$.
The first interesting quantities to calculate are the diffusion
exponent $\eta$ and the (anomalous) diffusion coefficient
$D_\eta(\tau)$ defined by means of the collision averaged mean square
displacement:

\begin{equation}
\label{eq-meansquaredis}
{\bf E}_{\mu,\tau,\sigma,T}((\vec{x}(t)-\vec{x}(0))^2)\;
\stackrel{{\textstyle \approx}}{{\scriptscriptstyle t\rightarrow
\infty}}
\;D_\eta(\tau)\;t^{2\eta}
\mbox{ , }
\end{equation}

\noindent where ${\bf E}_{\mu,\tau,\sigma,T}$ denotes the mean over
all collision times and outcomes. Clearly $\eta\in[0,1]$. In order to
give a precise mathematical meaning to the definition of $\eta$ and
$D_\eta(\tau)$, let us introduce the Laplace transform

\begin{equation}
\label{eq-laplace}
{\cal L}_{\mu,\tau,\sigma,T}(\delta)\;=\;\int^\infty_0dt\,e^{-\delta t}\,
{\bf E}_{\mu,\tau,\sigma,T}((\vec{x}(t)-\vec{x}(0))^2)
\mbox{ . }
\end{equation}

\noindent Then $\eta$ is defined to be half of the infimum over all
real $\gamma$ such that $\lim_{\delta\rightarrow 0}
\delta^{1+\gamma}{\cal L}_{\mu,\tau,\sigma,T}(\delta)=0$. An equivalent
definition is \cite{BeS}

\begin{equation}
\label{eq-diffexpodef}
\eta \; = \; \frac{1}{2}\,\limsup_{t\rightarrow \infty}\,
\frac{\mbox{Log}({\bf E}_{\mu,\tau,\sigma,T}
(\vec{x}(t)-\vec{x}(0))^2)}{\mbox{Log}(t)}
\mbox{ . }
\end{equation}

\noindent Whenever it exists, the diffusion coefficient is given by

\begin{eqnarray}
\label{eq-diffcoef}
D_\eta(\tau) & = &
\lim_{T\rightarrow \infty}
\int^T_0\frac{dt}{T}\;
\frac{{\bf
E}_{\mu,\tau,\sigma,T}((\vec{x}(t)-\vec{x}(0))^2)}{t^{2\eta}}
\nonumber
\\
& & 
\nonumber
\\
& = &
\lim_{\delta\rightarrow 0}
\delta^{1+2\eta}{\cal L}_{\mu,\tau,\sigma,T}(\delta)
\mbox{ , }
\end{eqnarray}

\noindent where the equality follows from a Tauberian lemma. The
Laplace transform ${\cal L}_{\mu,\tau,\sigma,T}$ can be calculated
explicitly:

\begin{equation}
{\cal L}_{\mu,\tau,\sigma,T}(\delta)\;=\;
\frac{2\sigma C_{T,\sigma}}{\delta}\;
\frac{\int^\infty_0 ds\, p_{\mu,\tau}(s)
\int^s_0dr\,e^{-\delta r}r^{2\sigma-1}}{1-
\int^\infty_0ds\,p_{\mu,\tau}(s)\,e^{-\delta s}}\;
\mbox{ . } 
\end{equation}

\noindent Hence the results can be summerized by the phase
diagram given in Figure 1 where $\eta$ and $D_\eta(\tau)$ in the regions
{\bf I} to {\bf IV} are given by
$$
\begin{array}{ccc}
\mbox{\bf I }\;\; & \eta=\frac{1}{2}\mbox{ , } 
& D_\eta(\tau)=\tau^{2\sigma-1}\,C_{T,\sigma}
\frac{\langle s^{2\sigma}\rangle_{\mu,1}}{\langle s\rangle_{\mu,1}}
\mbox{ , } 
\\
& & \\
\mbox{\bf II }\;\; & \eta=\sigma+\frac{1-\mu}{2}\mbox{ , } 
& D_\eta(\tau)=\tau^{\mu-1}\, C_{T,\sigma}
\,C^{\mbox{\tiny II}}_\mu\mbox{ , } 
\\
& & \\
\mbox{\bf III }\;\; & \eta=\mu \mbox{ , } 
& D_\eta(\tau)= \tau^{2\sigma-\mu}
\,C_{T,\sigma}
\,C^{\mbox{\tiny III}}_\mu\mbox{ , } 
\\
& & \\
\mbox{\bf IV }\;\; & \eta=\sigma
\mbox{ , } 
& D_\eta(\tau)= 
\,C_{T,\sigma}
\,C^{\mbox{\tiny IV}}_\mu
\mbox{ , }
\end{array}
$$

\noindent where $\langle \,.\,\rangle_{\mu,\tau}$ denotes the mean
with respect to $p_{\mu,\tau}$ and $C^{\mbox{\tiny II}}_\mu$,
$C^{\mbox{\tiny III}}_\mu$ and $C^{\mbox{\tiny IV}}_\mu$ are
numerical constants independent of $\tau$. Note that $\eta$ is well
defined and continuous on the critical lines separating the four
phases, because its definition is independent of subdominant
logarithmic terms. The diffusion coefficient as defined in
(\ref{eq-diffcoef}), however, does not exist on these lines because of
these logarithmic divergences. Nevertheless, the dependence of
$D_\eta$ on $\tau$ varies continuously across the critical lines. Now
follow comments on each of the four regions.

%%%%%%%%%%%%
\begin{figure}
\centerline{\psfig{figure=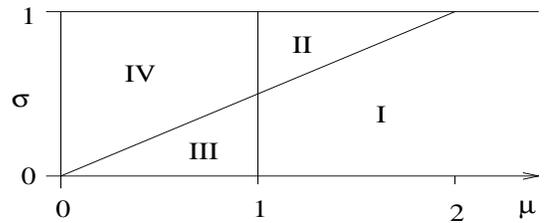,height=3cm,width=7cm,angle=270}}
\caption{{\sl  Phase diagram for the anomalous Drude model}}
\end{figure}
%%%%%%%%%%%%

The diffusive regime {\bf I} is remarkable because of the stability of
diffusion exponent and coefficient upon change of the probability law
for the time delays. The dependence of
$D_{1/2}(\tau)\sim\tau^{2\sigma-1}$ already appears in
\cite{SG,May,BeS}. Because $D_{1/2}(\tau)$ is linked to the
conductivity by an Einstein relation, this formula is of considerable
importance when studying electronic transport in quasicrystals. These
issues are further discussed below. In region {\bf II}, the collision
rate is not sufficient to force the collision averaged motion to be
diffusive and hence it stays anomalous superdiffusive ($\eta>1/2$). In
literature \cite{Man,BG}, the motion in this regime is generally
referred to as a L\'evy flight and, for $\sigma=1$, the above results
are known \cite{BG}. L\'evy flights can either be obtained from the
dynamics of a collisional model with long flying times (as above) or
alternatively with long jumps (as below) (for further references, see
\cite{BG,ZA}). The latter approach can be motivated by a maximal
entropy principle using Tsallis' generalized entropy \cite{ZA}.
Anomalous superdiffusion is of growing importance in various fields of
physics such as laser cooling \cite{BBE}, ionic transport in oxide
glasses \cite{SGB}, non-linear Hamiltonian dynamics \cite{ZSW}, among
others (see the review \cite{BG}). In region {\bf III}, the anomalous
motion is subdiffusive; the collision rate imposes the diffusion
exponent and only the diffusion coefficient depends on the free
motion. The case $\sigma=0$ is already studied in \cite{MW,MS}. In
region {\bf IV}, collisions are so rare that they do not affect the
anomalous free motion.

\vspace{.2cm}

The anomalous Drude model is equivalent to a continuous time random
walk \cite{MW,BG}. This model is defined by a probability
$\psi(\vec{r},s)ds$ that the walker (or particle) remains for a time
$s$ at a given site before performing a jump of length $\vec{r}$
between $s$ and $s+ds$. If
$\psi(\vec{r},s)=\hat{\psi}(\vec{r})\tilde{\psi}(s)$, the random walk
is said to be separable. For the anomalous Drude model, one supposes
that the particle stays at a given site according to
(\ref{eq-diffexpo}) and that the jump length is determined by
(\ref{eq-delaydist}) according to the length of the waiting time.
Hence the corresponding continuous time random walk is separable only
for $\sigma=0$.

It is interesting to compare the obtained results with a corresponding
discrete time random walk. In fact, as long as $\mu>1$ (regions {\bf
I} and {\bf II}), the collision  time $\langle s \rangle_{\mu,\tau}$
is finite. It is hence appealing to consider a random walker which
makes a jump at times  $N \langle s \rangle_{\mu,\tau}$, $N\in{\bf
N}$, with spherical symmetric jump probability 

\begin{equation}
\label{eq-probdist}
P_{\mu,\tau,\sigma,T}(\vec{x})d^dx\;=\;
p_{\mu,\tau}\left(
\frac{|\vec{x}|^{1/\sigma}}{C_{T,\sigma}^{1/2\sigma}}\right)
\,\frac{|\vec{x}|^{-1+1/\sigma}}{S^{d-1}C_{T,\sigma}^{1/2\sigma}}\;
d\Omega\,d|\vec{x}|
\mbox{ , }
\end{equation}

\noindent where $|\vec{x}|\geq 0$ is the radial variable in spherical
coordinates, $d\Omega$ the area element on the $(d-1)$-sphere and
$S^{d-1}$ the area of the latter. Note that
$P_{\mu,\tau,\sigma,T}(\vec{x})\sim |\vec{x}|^{-1-\mu/\sigma}$ as
$|\vec{x}|\rightarrow \infty$. The distribution of the position of the
walker after $N$ jumps is given by the $N$-th convolution product of
$P_{\mu,\tau,\sigma,T}$. The diffusion exponents $\eta'(q)$ and
anomalous diffusion coefficients $D_{\eta'(q)}'$, $q>0$, are then
introduced by

\begin{equation}
\label{eq-diffexpo2}
\int_{{\bf R}^d}d^dx\;P_{\mu,\tau,\sigma,T}^{*N}(\vec{x})\,
|\vec{x}|^q
\;\stackrel{{\textstyle \approx}}{{\scriptscriptstyle N\rightarrow
\infty}}
\;D_{\eta'(q)}'(\tau)\,N^{q\eta'(q)}
\mbox{ . }
\end{equation}
 
In region {\bf I}, $\mu>2\sigma$ so that the second moment of
$P_{\mu,\tau,\sigma,T}$ exists. Therefore the central limit theorem
implies that the $N$-th fold convolution product
$P_{\mu,\tau,\sigma,T}^{*N}(x_N)$ with $x_N=x/\sqrt{N}$ converges in
distribution to a Gaussian and one verifies that $\eta'(q)=1/2$
independently of $q$. Of course, this confirms the stability of this
regime addressed above. Moreover, the ratio of the second moment of
$P_{\mu,\tau,\sigma,T}$ and the collision time $\langle s
\rangle_{\mu,\tau}$ gives $D_{1/2}'(\tau)$. One has 
$D_{1/2}'(\tau)=D_{1/2}(\tau)$.

In region {\bf II}, $P_{\mu,\tau,\sigma,T}(x)\sim |x|^{-1-\gamma}$
with $\gamma=\mu/\sigma$ such that $1<\gamma<2$.  Hence Gnedenko's
generalized central limit theorem \cite{BG} implies that
$P_{\mu,\tau,\sigma,T}^{*N}(x_N)$  with the scaling
$x_N=x/{N}^{\gamma}$ converges in distribution to the spherical
symmetric L\'evy law $L_{\gamma,\tau,T}$ which is given by its
characteristic function

\begin{equation}
\label{eq-Levy}
\mbox{Log}(\hat{L}_{\gamma,\tau,T}(\vec{k}))\;=\;
{-C_{\gamma,T}\;|\tau\vec{k}|^\gamma}
\mbox{ , }
\end{equation}

\noindent where $C_{\gamma,T}$ is a constant depending on $\gamma$ and
$T$. Going through the details, this allows to determine the diffusion
exponents given by (\ref{eq-diffexpo2}) to be $\eta'(q)=\sigma/\mu$
for all $q<\mu/\sigma$. For $q>\mu/\sigma$ the integral in
(\ref{eq-diffexpo2}) diverges for any $N$. The anomalous diffusion
coefficients $D_{\eta'(q)}'(\tau)$ may be calculated as well.

At first sight, it seems to be paradoxical that $\eta'(q)$ is defined
only for $q<\mu/\sigma$ while $\eta$ is defined for $q=2$. In fact,
equation (\ref{eq-meansquaredis}) gives the stochastic evolution of the
mean square displacement. It exists for finite times because
$|\vec{x}(t)-\vec{x}(0)|\leq Ct^\sigma$ if the distribution of the
$\vec{v}_\sigma$ has a distribution supported in the ball of size $C$.
Therefore the diffusion exponent $\eta$ is well defined. On the other
hand, for the discrete time random walk with jump probability
$P_{\mu,\tau,\sigma,T}$, any step can have arbitrarily big length so
that the moment $q=2$ in (\ref{eq-diffexpo2}) diverges for any $N$. In
other words, the distribution of the particle positions after a
continuous time random walk of time $N\langle s\rangle_{\mu,\tau}$ is
compactly supported while that after $N$ steps distributed according to
(\ref{eq-probdist}) is slowly decaying. Hence there is no paradox at
all.

Note that nevertheless $\eta'(q)\leq\eta$ and their values
coincide only on the lines separating regime {\bf II} from the others.
This is due to the non-Markov property of the counting process
addressed above which favors long flying times and hence big mean
square displacement in (\ref{eq-meansquaredis}), but has no effect in
(\ref{eq-diffexpo2}).

\vspace{.2cm}

Let us now calculate the conductivity in the diffusive regime and
verify the Einstein relation. The accelerated particle motion in the
exterior electric field $\vec{\cal E}$ is

\begin{equation}
\label{eq-electric}
\vec{x}(t)\;=\;\vec{x}(0)+\vec{v}_\sigma t^\sigma
+q\vec{\cal E}\,c_{T,\sigma}\,t^{2\sigma}
\mbox{ , }
\end{equation}

\noindent where $q$ is the charge of particle and $c_{T,\sigma}$ is a
constant depending on $\sigma$ and temperature. As
(\ref{eq-diffexpo}), equation (\ref{eq-electric}) does not describe a
classical Hamiltonian motion, but should be seen as a modelization of
the quantum motion in an exterior electric field. In fact, if the
Hamiltonian quantum dynamics without electric field leads to
(\ref{eq-diffexpo}), then the motion in presence of an electric field
satisfies (\ref{eq-electric}) in linear approximation in $\vec{\cal
E}$ \cite{futwork}. For ballistic free motion, (\ref{eq-electric})
gives the usual accelerated motion. If $\sigma=0$, (\ref{eq-electric})
means that the hopping between localized states is preferably in
direction of the electric field. For $\sigma=1/2$, the motion
(\ref{eq-electric}) is ballistic in presence of an electric field
which allows to calculate the conductivity even if there are no
collisions. In any other case ($\sigma\neq 1/2$), collisions are
necessary as impulsion and energy sink (for $\sigma>1/2$) or source
(for $\sigma<1/2$). Strictly speaking, energy dissipation is only
needed when studying thermal transport (recall that the electrical 
current can be calculated in the classical Lorentz model although 
collisions in this model are elastic).

The average speed $\langle \vec{v}\rangle_{\vec{{\cal
E}},\sigma,\mu,\tau,T}$ (in the usual sense) of a particle evolving
between collisions with (\ref{eq-electric}) is given by the long time
average of ${\bf E}_{\mu,\tau,\sigma,T}(\vec{x}(t)-\vec{x}(0))/t$ or
equivalently by

\begin{equation}
\label{eq-meanspeed}
\langle \vec{v}\rangle_{\vec{{\cal E}},\sigma,\mu,\tau,T}
\,=\;
\lim_{\delta\rightarrow 0}
\delta^2\!\int^\infty_0\!dt\;e^{-\delta t}
{\bf E}_{\mu,\tau,\sigma,T}(\vec{x}(t)-\vec{x}(0))
\mbox{ . }
\end{equation}

\noindent For this limit to exist and to be finite, parameters have
to be in the diffusive regime {\bf I}. The calculation gives $\langle
\vec{v}\rangle_{\vec{{\cal E}},\sigma,\mu,\tau,T}=q
c_{T,\sigma}D_{1/2}(\tau)\vec{{\cal E}}/C_{T,\sigma}$. If $n$ is the
particle density, the direct conductivity is thus

\begin{equation}
\label{eq-conduct}
\hat{\sigma}\;=\;\tau^{2\sigma-1}{q^2n\,c_{T,\sigma}}
\frac{\langle s^{2\sigma}\rangle_{\mu,1}}{\langle s\rangle_{\mu,1}}
\mbox{ , }
\end{equation}

\noindent and the Einstein relation reads

\begin{equation}
\hat{\sigma}\;=\;
\frac{q^2n\,c_{T,\sigma}}{C_{T,\sigma}}
\,D_{1/2}(\tau)
\mbox{ . }
\end{equation}

\noindent The {\sl anomalous Drude formula} (\ref{eq-conduct}) was
obtained in \cite{SG,May} and later on in a purely quantum mechanical
context in \cite{BeS,futwork}.

Using the physical origin of the exponent $\sigma$ as indicated above,
the interpretation of (\ref{eq-conduct}) is the following. If
$\sigma<1/2$, the particles are trapped by the quantum interference
phenomena in the on-site potential and, in order to have appreciable
conductivity, there is a need for non-elastic collisions (as in Mott's
hopping conductivity). Consequently the conductivity increases with
increasing collision rate (smaller $\tau$) and vanishes if there are
no collisions ($\tau=\infty$). On the other hand, if $\sigma>1/2$, the
conductivity is infinite if there are no collision and it decreases
with increasing collision rate because the collisions slow down the
free superdiffusive motion. If $\sigma=1/2$, the conductivity is
independent of the collision rate.

%%%%%%%%%%%%
\begin{figure}
\centerline{\psfig{figure=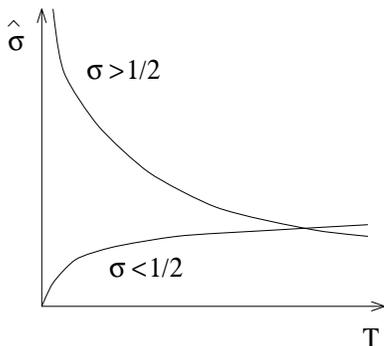,height=4.5cm,width=5cm,angle=270}}
\caption{{\sl  Schematic representation of the temperature behavior of
the conductivity as given by the anomalous Drude formula}}
\end{figure}
%%%%%%%%%%%%

\vspace{.2cm}

The mechanism just described is probably the physical origin for the
Mooij plot of quasicrystals with differing quality. Experiment shows
\cite{Ber}, that for those with a conductivity bigger than some
critical value, the temperature derivative of the conductivity at low
temperature  is negative, and for those with a conductivity smaller
than the critical value, the derivative is positive. For the
phenomenological explanation, one supposes that $\tau\sim T^{-\alpha}$
for some $\alpha>0$. According to (\ref{eq-conduct}), this leads to
the behavior of the conductivity given in Figure 2. Now, the higher
the quality of the quasicrystal, the lower is the diffusion exponent
(this expresses the continuous metal-insulator transition in
quasicrystals). The combination of these two facts explains the Mooij
plot; the critical value corresponds to the materials with
$\sigma=1/2$. Other phenomena such as the inverse Mathiessen rule can
be qualitatively explained by the anomalous Drude formula
\cite{futwork}.

\vspace{.2cm}

In summary, the anomalous Drude model modelizes the  quantum motion of
a particle in an on-site potential which further undergoes collisions
with phonons or other particles by a stochastic process. The diffusion
exponent and coefficient of the model show complex scaling laws.
Furthermore the electrical conductivity can be calculated in the
diffusive regime and the Einstein relation is valid. 

\vspace{.2cm}

Acknowledgement: the author would like to thank J. Bellissard, C. Sire,
M. Zarraouti and R. Fleckinger for numerous discusions and comments.

\end{document}